\documentclass[12pt,preprint]{aastex}

\shorttitle{Variability of Sgr A* at 3-mm}
 \shortauthors{Li et al.}

\begin{document}

\title{The variability of Sagittarius A* at 3 millimeter}


\author{Juan Li\altaffilmark{1,2}, Zhi-Qiang Shen\altaffilmark{1}, Atsushi Miyazaki\altaffilmark{3},
Lei Huang\altaffilmark{4}, R. J. Sault\altaffilmark{5}, Makoto
Miyoshi\altaffilmark{3}, Masato Tsuboi\altaffilmark{6} and Takahiro
Tsutsumi\altaffilmark{3}}

\altaffiltext{1}{Key Laboratory for Research in Galaxies and
Cosmology, Shanghai Astronomical Observatory, Chinese Academy of
Sciences, 80 Nandan RD, Shanghai 200030, China; lijuan@shao.ac.cn,
zshen@shao.ac.cn}
 \altaffiltext{2}{Graduate School of the Chinese
Academy of Sciences, Beijing 100039, China}
 \altaffiltext{3}{National Astronomical
Observatory of Japan, 2-21-1 Osawa, Mitaka, Tokyo 181-8588, Japan;
amiya@miz.nao.ac.jp, makoto.miyoshi@nao.ac.jp, t.tsutsumi@nao.ac.jp}
\altaffiltext{4}{Key Laboratory for Research in Galaxies and
Cosmology, University of Science and Technology of China, Hefei
230026, China; mlhuang@ustc.edu.cn}
 \altaffiltext{5}{University of Melbourne, School of Physics, Parkville,
Victoria 3052, Australia; rsault@unimelb.edu.au}
\altaffiltext{6}{Institute of Space and Astronautical Science,
Sagamihara, Kanagawa 229-8510, Japan; tsuboi@vsop.isas.jaxa.jp}


\begin{abstract}

We have performed monitoring observations of the 3-mm flux density
toward the Galactic Center compact radio source Sgr A* with the
Australia Telescope Compact Array since 2005 October. Careful
calibrations of both elevation-dependent and time-dependent gains
have enabled us to establish the variability behavior of Sgr A*.
Sgr~A* appeared to undergo a high and stable state in 2006 June
session, and a low and variable state in 2006 August session. We
report the results, with emphasis on two detected intra-day
variation events during its low states. One is on 2006 August 12
when Sgr~A* exhibited a 33\% fractional variation in about 2.5 hr.
The other is on 2006 August 13 when two peaks separated by about 4
hr, with a maximum variation of 21\% within 2 hr, were seen. The
observed short timescale variations are discussed in light of two
possible scenarios, i.e., the expanding plasmon model and the
sub-Keplerian orbiting hot spot model. The fitting results indicate
that for the adiabatically expanding plasmon model, the synchrotron
cooling can not be ignored, and a minimum mass-loss rate of $9.7
\times10^{-10}M_{\odot}$ yr$^{-1}$ is obtained based on parameters
derived for this modified expanding plasmon model. Simultaneous
multi-wavelength observation is crucial to our understanding the
physical origin of rapid radio variability in Sgr A*.

\end{abstract}


\section{Introduction}
\qquad

There is compelling evidence that Sagittarius A$^{\ast}$ (Sgr A*),
the extremely compact radio source at the dynamical center of the
Galaxy, is associated with a $4\times 10^{6}{\rm M}_\odot$ black
hole (Eckart \& Genzel 1996; Ghez et al. 2000; Sch$\ddot{\rm o}$del
et al. 2002; Eisenhauer 2003). Since its discovery in 1974 (Balick
\& Brown 1974), Sgr A* has been observed extensively with radio
telescopes in the northern hemisphere, and temporal flux variations
at millimeter wavelengths were reported. With VLA observations,
Yusef-Zadeh et al. (2006b) detected an increase of flux density at a
fractional level of 7\% and 4.5\% at 7- and 13-mm, respectively,
with a duration of about 2 hr. The peak flare emission at 7-mm led
the 13-mm peak flare by 20-40 minutes. Mauerhan et al. (2005)
detected intra-day variations (IDVs) of about 20\% and in some cases
up to 40\% at 3-mm using the Owens Valley Radio Observatory (OVRO).
The rise and decay occurred on a timescale of 1-2 hr. At 2-mm,
Miyazaki et al. (2004) reported a 30\% flux increase in 30 minutes
from the monitoring of the Nobeyama Millimeter Array (NMA). On the
other hand, flares with violent intensity increases in very short
timescales have also been detected at infrared and X-ray bands
(Genzel et al. 2003; Baganoff et al. 2001; Eckart et al. 2006b),
inferring that these emissions from Sgr A* originate within very
vicinity of the central massive black hole. This is further
strengthened by the simultaneous detection of X-ray, infrared and
sub-mm flares (Eckart et al. 2004, 2006a, 2008a, 2008b; Yusef-Zadeh
et al. 2006a, 2008; Marrone et al. 2008).

Since Sgr A* is embedded in thick thermal material, it is
particularly difficult to observe its intrinsic structure. But
observations of IDV can give indirect constraints on the source
emission geometry and emission mechanisms. However, previous
monitoring observations of Sgr A* from the northern hemisphere have
been strictly limited to a short observing window ($<$ 7~hr/day) for
the Galactic Center region. We have performed monitoring
observations of flux density toward Sgr A* at 3-mm since 2005
October when for the first time the Australia Telescope Compact
Array (ATCA) of the Australia Telescope National Facility (ATNF) was
available at 3-mm. The ATCA is an interferometer consisting of five
22-m radio telescopes at Narrabri, Australia where Sgr A* passes
almost overhead, allowing a much longer observing window ($>$ 8~hr
at elevation angles above 40$^{\circ}$). As such, the ATCA
calibrations and flux density measurements of Sgr A* are expected to
be more accurate.

In this paper, we report our effort to search for IDV in Sgr A* with
the ATCA. We first introduce the ATCA observations in \S\
\ref{observation}. The data reduction and analysis with emphasis on
the gain calibrations are described in detail in \S\ \ref{analysis}.
In \S\ \ref{result}, we present the detection of IDV events in
Sgr~A*. To interpret the observation, we discuss two possible scenarios in \S\
\ref{discussion}, followed by a summary in \S\ \ref{summary}.
Throughout this paper, the fractional variation is defined as
$\frac{S_{max}-S_{min}}{\frac{1}{2}(S_{max}+S_{min})}$, here
$S_{max}$ and $S_{min}$ refer to the maximum and minimum value of
flux density, respectively.

\section{OBSERVATIONS}
\label{observation}

In 2005 and 2006, we performed 3-mm ATCA flux density monitoring of
Sgr~A* over 50 hr in the following 3 sessions: 2005 October 18, 2006
June 9 and August 9-13.
 Dual (linear) polarization double sideband (DSB)
HEMT receivers were used. The first ever 3-mm ATCA monitoring of
Sgr~A* was performed on 2005 October 18 when the data were
simultaneously recorded, in both the lower (93.504 GHz) and upper
(95.552 GHz) sidebands, in 32 channels of a total bandwidth of 128
MHz. For the observations in 2006, the data were recorded in two
slightly different 3-mm bands: the lower sideband (86.243 GHz) was
set to the transition frequency of the SiO J=2-1 v=1 line with 256
channels of a total bandwidth of 16 MHz, the upper sideband (88.896
GHz) was a wideband with 32 channels of a total bandwidth of 128
MHz. Since the continuum data of the lower sideband with narrow
bandwidth have relatively low signal to noise ratio, only the upper
sideband data were used for Sgr A* and other continuum sources.

On 2005 October 18, we observed Sgr A* in the H168C configuration of
the ATCA, with a maximum baseline of 192~m, uv range of 13 -
61k$\lambda$ and a synthesized beam of $2.^{\prime\prime}9 \times
1.^{\prime\prime}7$. On 2006 June 9, the observations were performed
in the 1.5D configuration with a maximum baseline of 1439~m,
covering uv range of 20 - 430k$\lambda$ and yielding a synthesized
beam of $2.^{\prime\prime}1 \times 0.^{\prime\prime}3$. In 2006
August, the observations were performed in SPLIT5 configuration with
a maximum baseline of 1929~m, covering uv range of 3 - 570k$\lambda$
and yielding a synthesized beam of $1.^{\prime\prime}3 \times
0.^{\prime\prime}2$. In this array, the spacing between antennas 2
and 3 and antennas 3 and 4 are only 31~m, causing severe shadow
effect, especially for antenna 3.

Quasar 3C 279 was observed for 10 minutes at the beginning of the
observation to calibrate bandpass. Either a planet or a
bright radio source was observed for the flux density calibration.
The first 3-mm ATCA observation of Sgr A* lasted for 10 hr on 2005
October 18. It alternated between Sgr A* and the only secondary
calibrator PKS~1730-130, which was also used as pointing calibrator.
The primary calibrator Uranus was observed for 10 minutes at the end
of the observation. In 2006 June and August, we observed Sgr A* in a
total of 6 days with a thoughtful calibration strategy. Up to four
secondary calibrators (control sources) including an SiO maser
source (OH2.6-0.4) and three continuum sources (PKS 1921-293, PKS
1710-269 and PKS 1730-130) were observed to check the consistency of
the gain calibrations.
Limited by the weather, ATCA observed Sgr A* for only 1 hr on
 August 9, 4 hr on August 10 and 2 hr on August 11. Therefore, we
 will focus on the measurements on June 9, August 12 and 13.
The pointing accuracy was checked  every half an hour by observing
VX Sgr, a known strong SiO maser source. The instrumental gain and
phase were calibrated by alternating observations of Sgr A* and all
secondary calibrators. In 2006 June, the observations were performed
using the following sequence: OH2.6-0.4 (2 min), Sgr A* (5 min), PKS
1730-130 (2 min), and PKS 1921-293 (1 min). In 2006 August, the
observing sequence was PKS 1710-269 (2 min), OH2.6-0.4 (1 min), Sgr
A* (10 min),
 PKS 1710-269 (2 min), OH2.6-0.4 (1 min), PKS
1730-130 (1 min),  and PKS 1921-293 (1 min). The observing details
have been summarized in Table 1.

\section{DATA REDUCTION AND ANALYSIS}
\label{analysis}

All the data processing was conducted using the ATNF MIRIAD package
(Sault et al. 1995). At millimeter wavelengths, the atmosphere can
no longer be approximately transparent. The opacity effect is
included in an effective system temperature - the so-called ``above
atmosphere" system temperature (Ulich 1980) for the ATCA
measurements at 3-mm. The bandpass corrections were made using the
strong ATCA calibrator 3C279. For amplitude calibration, we first
applied a nominal elevation-dependent gains of the antennas and then
used calibrators to further determine the additional corrections. On
2005 October 18, the flux scale was based on observation
 of Uranus. On 2006 June 9, the flux density scale was determined
 using PKS~1730-130, assuming its flux density of 2.27~Jy at 3-mm. In
 2006 August, we derived the flux density scale with another
brighter radio source PKS~1921-293, which is reported to be 8.44~Jy
during our observations from the ATCA calibrator list on web. From
the ATCA calibrator flux density monitoring data during 2003 to
2006, we estimated its mean flux density of 8.66 Jy with a standard
deviation of 1.04, implying a dispersion of about 12\%. PKS 1921-293
is probably better than that of other calibrators simply because PKS
1921-293 data usually have very high signal-to-noise ratio. So, we
expect an accuracy $\leq$ 20\% for the absolute amplitude
calibration in these observations. After the phase self-calibration,
the data were averaged in 5 minutes bin to search for shorter
timescale variability. The flux density of Sgr A* was estimated by
fitting a point source model to visibilities on the projected
baselines longer than 25k$\lambda$ (about 85 m at 3-mm) to suppress
the contamination from the surrounding extended components (Miyazaki
et al. 2004; Mauerhan et al. 2005). Both the fitting error reported
by MIRIAD and the rms of the residual visibilities were used to get
the final error estimate.

In order to establish strong cases for variability of Sgr A* at
millimeter wavelengths, reliable calibrations of both elevation- and
time-dependent gains are crucial. We have carefully considered and
corrected the following factors that could affect the measurements
during the calibration process.

1. Antenna gain varies with elevation angle mainly because of the
gravitational distortion of the dish. The antenna efficiency of ATCA
has maximum value at an elevation angle of $60^{\circ}$ and minimum
value at an elevation angle of $90^{\circ}$. A nominal
gain-elevation correction in MIRIAD is applied at elevations greater
than $40^{\circ}$ for 3-mm observations, only those data observed at
elevation angles above $40^{\circ}$ were used. However, such nominal
elevation-dependent gains built in MIRIAD seem hard to fully
compensate the gain variation. We have plotted flux density as a
function of elevation angle and found that a nearby calibrator was
needed to make further correction, otherwise significant elevation
effect (e.g., peaks at about $60^{\circ}$, or reaches the lowest
point at $90^{\circ}$ elevation angle) will be shown up in the
light-curve, which often indicates some calibration errors.
PKS~1730-130, which is often used to calibrate phase and amplitude
during observations from northern hemisphere at mm wavelength (e.g.
Miyazaki et al. 2004, Yusef-Zadeh et al. 2008), is proved to be
unsuitable for the ATCA observations. It is 16.2$^\circ$ away from
Sgr~A*, and its elevation angle is only 72$^\circ$ when Sgr A*
reaches the zenith. Thus, gain corrections derived from this source
data cannot fully compensate the elevation effect in Sgr A*,
especially for observations at high elevations. Similarly,
PKS~1921-293, which reaches the zenith 2 hr later than Sgr A*, is
not suitable, either. So we only use two closer sources PKS~1741-312
and OH2.6-0.4 for the gain calibration.

2. Calibrators are, in general, variable sources which will
unavoidably introduce uncertainties into the nominal
time-independent gains. For this reason, several secondary
calibrators were actually scheduled to check the consistency. The
complex gains derived from one control source were applied to both
Sgr A* and other control sources. If such a control source is
strongly variable, a somehow similar trend in light-curve will
appear for all the other sources (including Sgr A*) after
calibration.

To check the significance of any detected variability, we introduced
the modulation index, which is defined as the rms of the gain
correction of five antennas derived from calibrators and flux
density of Sgr A* divided by their mean, corresponding to the degree
of variation for Sgr~A*, and the fractional uncertainty in
time-dependent gain correction, respectively. Obviously, if the
modulation index of Sgr~A* flux density is much larger than that of
antenna gain correction, the detected flux variation is most likely
to be real. The modulation indices of Sgr A* and gain correction of
five antennas derived from two nearby calibrators OH2.6-0.4 and
PKS~1710-269 on 2006 June 9, August 12 and 13 are plotted in Figure
1. The modulation indices of Sgr A* were quite large on 2006 August
12 and 13, indicating the real detection of IDV from Sgr~A*. As
mentioned in \S\ \ref{observation}, many data obtained from antenna
3 in 2006 August were shadowed and thus not used in obtaining its
gain correction, resulting in a large fluctuation in its gain
correction and thus a bigger modulation index. During observations
in 2006 August, the 3-mm flux density of PKS~1710-269 is around
0.5~Jy, only one fiftieth of that of OH2.6-0.4, therefore the
signal-to-noise ratio of PKS 1710-269 data is much lower than that
of OH2.6-0.4. This explains why the modulation indices of gain
corrections derived from PKS~1710-269 are relatively high.

3. As mentioned in \S\ \ref{observation}, we only used upper
sideband (88.896 GHz) data with a bandwidth of 128 MHz for Sgr A*
and other continuum control sources, and the lower sideband (86.243
GHz) data of 32 MHz bandwidth only for the SiO maser source
OH2.6-0.4. Will there be an additional uncertainty when applying to
Sgr A* the gain solutions derived from OH2.6-0.4 data? We inspect
this by comparing the results of the two sidebands on 2006 August 12
and 13. Similar to what we did for the upper sideband data, the flux
densities of Sgr~A* was also estimated from the lower sideband using
the same channels as OH2.6-0.4. The results from the lower sideband
data show larger error bars mainly because of the relatively low
signal-to-noise ratio. The average deviations from results of upper
sideband data are 2.4\% on August 12 and 3.6\% on August 13, much
smaller than the fractional variation of Sgr A* (see \S\
\ref{result}).

4. We also consider the response of feeds to polarized emission. The
feed of ATCA is linearly polarized, and its response to a signal is
a combination of total and linear polarized intensity. Thus, two
polarization products, XX and YY correlations can be used as a
direct measure of total intensity only when a calibrator is not
linearly polarized. Unfortunately, polarizations of nearly 27\% were
observed in OH2.6-0.4 (Glenn et al. 2003). In addition, for ATCA
antennas on altazimuth mounts, their feeds rotate with respect to
the equatorial frame. This causes the actual response of ideal
linearly polarized feeds to vary with the parallactic angle. To
solve this problem, we used total intensity by summing the two
polarization products (XX and YY) to remove the effect of linear
polarization of the SiO maser OH2.6-0.4, and then derived the gain
correction. In other words, a single joint solution was determined.
This is generally a reasonable approximation given that antenna
gains are dominated by changes common to both polarizations and, the
difference between them is only a few percent and can be ignored
safely (Maxim Voronkov, private communication).

Overall, OH2.6-0.4, which is relatively stable and close to Sgr A*
(about 2.7$^\circ$ away), proved to be the best control source. As
such, it was used as the main secondary calibrator to determine the
antenna gain corrections for all the results presented in this
paper.

\section{RESULTS}
\label{result}

During the first 3-mm ATCA observation of Sgr A* on 2005 October 18,
the flux density of Sgr A* ran up to 3.5 Jy, much brighter than the
normally expected $1.5$~Jy in the quiescent phase, and thus Sgr A*
was very likely to be in an active phase during our observation.
Although Sgr A* seems to vary in its total flux density as a
function of time, we can not rule out the possibility of the
elevation-dependent gain effect as use of the only secondary
calibrator PKS~1730-130 (16.2$^\circ$ from Sgr~A*) severely limited
the amplitude calibration. Because of this, starting from
observations in 2006 June and August (see \S\ \ref{observation}), we
paid a particular attention to the strategy of calibration. As a
result, the light-curves of Sgr A* at 3-mm in 2006 are reliably
obtained (Figure 2). All the data were calibrated using OH2.6-0.4.
The flux densities were estimated by fitting a point source model to
visibility data on the projected baselines longer than 25k$\lambda$.
Following is a detailed description of the results from each
observation.

The flux density of Sgr A* was relatively high (around 3 Jy) but
stable on 2006 June 9. As shown in Figure 1, the modulation index of
Sgr~A* is small and comparable to that of antenna gains. Therefore,
no IDV was detected.

During the first three days in the 2006 August session (August 9, 10
and 11), very limited data were available. The flux density of
Sgr~A* was decreased from 2.52 to 2.25 Jy in 1 hr on August 9,
stayed around 1.9 Jy quite stably during the 4 hr run on August 10
and around 2.0 Jy over the 2 hr observation on August 11. So, we
conclude that no ascertained IDV was detected.

Two clear IDV events were seen in the last two days of the 2006
August session. As shown in the light-curves of Sgr A* and other
sources on 2006 August 12 (Figure 3 left), first the flux density of
Sgr~A* decreased from 1.65 to 1.50 Jy, and then increased to 2.11 Jy
in 2.5 hr before decreasing again to 1.90 Jy. The fractional flux
density variation is estimated to be $33\%$. On 2006 August 13
(Figure 3 right), the flux density of Sgr A* first increased from
1.95 to 2.14 Jy, reached its first peak before decreasing to 1.80 Jy
in 1.7 hr. Then it reached the second peak 2.22 Jy in 1.9 hr, and
declined to 1.98 Jy in 1.2 hr. The maximum fractional flux density
variation is $21\%$ with a timescale of about 2 hr. As shown in
Figure 1, the modulation indices of Sgr A* on both August 12 and 13
are much greater than that of gain corrections, supporting that the
observed flux density variations are most likely to be real.

The NMA observations of Sgr A* from 1996 to 2003 indicate that Sgr
A* has quiescent and active phases, the peaks of flares were 2-3 Jy
at 3-mm while the mean flux density in a quiescent phase was 1.1
$\pm$ 0.2 Jy at 90 GHz (Tsutsumi et al. 2002, Miyazaki et al. 2003),
which are in accord with our ATCA observations. As is shown in
Figure 2, the mean flux density of Sgr A* dropped from 2.97 to 2.16
Jy from 2006 June to August session. The day-to-day fractional
variation of Sgr A* appeared to be low from 2006 August 10 to 13.
Comparison of flux densities in two observing sessions in 2006
indicates that Sgr A* appeared to undergo a high state in 2006 June
session, and a low state in 2006 August session. Such different
states were also noted by Herrnstein et al. (2004). They found a
bimodal distribution of flux densities at centimeter wavelength and
thought that it might indicate the existence of two distinct states
of accretion onto the supermassive black hole. Different radiation
states usually have connections to certain physical parameters or
radiation model. For example, a unified inner advection dominated
accretion flow (ADAF) model with different accretion rates and
consequently different geometries of accretion flow has been
proposed to explain five distinct spectral states that have been
identified in black hole X-ray binaries (BHXBs), namely the
quiescent, low, intermediate, high and very high states (Esin et al.
1997). Supposing that the accretion model of Sgr A* has something in
common with that of BHXBs, the accretion rate in 2006 August session
should be smaller than that in June session, but the accretion rate
may not change much over days in August.

\section{DISCUSSIONS}
\label{discussion}

Several models have been invoked to explain the flaring activity of
Sgr A*, such as the expanding plasmon model and orbiting hot spot
model. We will discuss them separately.

\subsection{The Plasmon Model}

Expanding plasmon model of van der Laan (1966) was invoked to
explain observed time delay in variation of Sgr A* at 7- and 13-mm
(Yusef-Zadeh et al. 2006a, 2006b, 2008). In this model, rather than
the synchrotron cooling, the adiabatic cooling associated with
expansion of the emitting plasma is responsible for the decline of
flare. Flaring at a given frequency is produced through the
adiabatic expansion of an initially optically thick blob of
synchrotron-emitting relativistic electrons. The initial rise of the
flux density is produced by the increase in the surface area of blob
while it still remains optically thick; the curve turns over once
the blob becomes optically thin because of the reduction in the
magnetic field, the adiabatic cooling of electrons, and the reduced
column density as the blob expands. Such kind of blob ejected from
an ADAF
 is also thought to be a possible explanation for nonthermal flares and
 recombination X-ray lines in low-luminosity active galactic nuclei and
 radio-loud quasars (Wang et al. 2000).

Our observed IDVs with different amplitudes and timescales seem
consistent with the expanding plasmon model in the context of jet or
outflow. The amplitudes and timescales vary with the relativistic
particle energy distribution, expanding velocity and size of the
blob. To apply the model to the light-curves on 2006 August 12 and
13, we first assumed a power-law spectrum of the relativistic
particle energy ($n(E)\propto E^{-p}$). Hornstein et al. (2007)
reported a constant spectral index of 0.6 using multi-band IR
observations of several flares. Here we adopt a spectral index of
0.6, corresponding to the particle spectral index of 2.2, the energy
of the particles was assumed to range from 10 MeV to 3 GeV. The
expanding velocity was supposed to be constant. As is stated by
Yusef-Zadeh (2008), the relationship between the quiescent and
flaring states of Sgr A* is not fully understood. Their results
indicate that the quiescent emission at 7- and 13-mm varies on
different days. The minimum flux density was 1.5 Jy during our 2006
August observing session, the quiescent flux density, if it does
exist, should not be more than this value. We then assume a
quiescent flux density of 1.4 Jy, while the flare is produced by the
blob. Other parameters were derived by means of the weighted least
square method. We adopt exponentially increasing step length for
number density during the fitting in order to improve efficiency.
The uncertainties of the parameters were assessed by scaling up the
68.3 \% confidence region of parameter space, as an increase of
$\chi^{2}$ from $\chi^{2}_{min}$ to $\chi^{2}_{min} + \chi^{2}_{\nu}
$ with the reduced chi squares,
$\chi^{2}_{\nu}=\chi^{2}_{min}/N_{dof}$, where $N_{dof}$ is the
difference between the number of data and the number of fitting
parameters (c.f. Shen et al. 2003).

We used two blobs to fit for flare observed on 2006 August 12 and
three blobs for those observed on 2006 August 13. Initial magnetic
field of 20-50 Gauss were derived from the fit. The electron cooling
timescale due to synchrotron loss is (e.g., Marrone et al. 2008)
\begin{equation}
    t_{\rm syn} = 38 \left(\frac{\nu}{90}\right)^{-1/2}\left(\frac{B}{10}\right)^{-3/2} [\rm hr].
\end{equation}
where the frequency ($\nu$) is in GHz and magnetic field (B) in
Gauss. It is about 3.4 hr with a magnetic field of 50 Gauss at 90
GHz, which is comparable to the observed decreasing timescale of 2
hr. Thus the synchrotron cooling of the electrons should not be
ignored. We took this into account and re-did the whole fit. The
energy loss rate is given by You (1998):
\begin{equation}
    (\frac{d \gamma}{dt})_{\rm syn} = -3\times 10^{-8} \gamma^{2}U_{\rm mag},
\end{equation}
where $U_{\rm mag}=\frac{B^{2}}{8\pi}$. With a constant expanding
velocity $v$, the radius of the blob $R$ can be expressed as
$R=R_{0}+vt$, $R_0$ is the initial radius of the blob at a specific
instant $t_0=0$. Substituting $B$ and $t$ with $v$, $R$, $R_{0}$ and
the initial magnetic field $B_0$, Eq.(2) can be written as
\begin{equation}
    (\frac{d \gamma}{dR})_{\rm syn} =\frac{1}{v}(\frac{d \gamma}{dt})_{\rm syn} =-\frac{3\times
    10^{-8}B_0^2R_0^{4}}{8\pi v}\gamma^{2}R^{-4}=-c_1\gamma^2R^{-4}
\end{equation}
where $c_1=\frac{3\times 10^{-8}B_0^2R_0^{4}}{8\pi v}$. The energy
loss rate due to the adiabatical expanding is
\begin{equation}
(\frac{d\gamma}{dR})_{\rm exp}=-\frac{\gamma}{R}.
\end{equation}
Thus the total energy loss rate due to both synchrotron cooling and
expanding is
\begin{equation}
\frac{d\gamma}{dR}=-\frac{\gamma}{R}-c_1\gamma^2R^{-4}
\end{equation}
Eq. (5) is a Bernoulli equation with a solution
\begin{equation}
\gamma=\gamma_0 \left(\frac{R}{R_0}\right)^{-1}\left\{\frac{1}{4}c_1
\gamma_0 R_0^{-3}
\left[1-\left(\frac{R}{R_0}\right)^{-4}\right]+1\right\}^{-1}.
\end{equation}
Then the optical depth scales as
\begin{equation}
    \tau(\nu,R) = \tau(\nu_0,R_0) \left(\frac{\nu}{\nu_0}\right)^{-(p+4)/2}
    \left(\frac{R}{R_0}\right)^{-(2p+3)}\left\{\frac{1}{4}c_1
\gamma_0 R_0^{-3}
\left[1-\left(\frac{R}{R_0}\right)^{-4}\right]+1\right\} ^{1-p}
    \label{eq:tau}
\end{equation}
and the flux density scales as
\begin{equation}
        S(\nu,R) = S(\nu_0,R_0) \left(\frac{\nu}{\nu_0}\right)^{5/2}
    \left(\frac{R}{R_0}\right)^3
    \frac{1-\exp(-\tau(\nu,R))}{1-\exp(-\tau(\nu_0,R_0))}.
\end{equation}
where $\tau(\nu_0,R_0)$, $S(\nu_0,R_0)$ are optical depth and flux
density for frequency $\nu_0$ at the specific instant $t_0$. The
critical optical depth $\tau_{crit}(R)$, at which the flux density
for any particular frequency peaks for radius $R$, satisfies
\begin{equation}
e^{\tau_{crit}(R)}-\frac{1}{3}(2p+3)\tau_{crit}(R) - C_2(R)
\tau_{crit}(R)-1=0
\end{equation}
where $C_2(R)=\frac{1}{3}c_1 \gamma_0 R_0^{-3}(p-1)
\left(\frac{R}{R_0}\right)^{(-4)} \left\{\frac{1}{4}c_1 \gamma_0
R_0^{-3} \left[1-\left(\frac{R}{R_0}\right)^{-4}\right]+1\right\}
^{-1}$. In the expanding plasmon model of van der Laan (1966),
optical depth scales as
\begin{equation}
    \tau(\nu,R) = \tau(\nu_0,R_0) \left(\frac{\nu}{\nu_0}\right)^{-(p+4)/2}
    \left(\frac{R}{R_0}\right)^{-(2p+3)},
    \label{eq:tau}
\end{equation}
 and $\tau_{crit}(R)$, the critical optical depth at the maximum of the light
curve at any frequency, depending only on p through the equation
(Yusef-Zadeh et al. 2006b, 2008)
\begin{equation}
e^{\tau_{crit}(R)}-(2p/3+1)\tau_{crit}(R)-1=0.
\end{equation}
Comparison between Eq.(7) and (10) indicates that the only
difference between these two equations is factor
$\left\{\frac{1}{4}c_1 \gamma_0 R_0^{-3}
\left[1-\left(\frac{R}{R_0}\right)^{-4}\right]+1 \right\}^{1-p}$ in
the former, which is the result of synchrotron cooling. Similarly,
the only difference between Eq.(9) and (11) is factor $C_2(R)$ of
$\tau_{crit}(R)$ in the former, which decreases as $R^{-4}$. The
optical depth at which the flux density peaks at $t_0$ satisfies
\begin{equation}
e^{\tau_{crit}(R_0)}-\frac{1}{3}(2p+3)\tau_{crit}(R_0) -
C_2(R_0)\tau_{crit}(R_0)-1=0.
\end{equation}
For typical values of $p=2.2$, $B_0=20$ Gauss, $\gamma_0=20$,
$R_0=4r_g$ and $v=0.004c$,
\begin{equation}
 C_2(R_0)=\frac{1}{3}c_1 \gamma_0 R_0^{-3}(p-1)=\frac{1\times 10^{-8}B_0^2 \gamma_0 R_0
(p-1)}{8\pi v}=0.08 \ll \frac{1}{3}(2p+3)=2.5
\end{equation}
which implies that $\tau_{crit}(R_0)$ mainly depends on $p$. Since
$c_1 \gamma_0 R_0^{-3}(p-1) \propto B_0^2$, $B_0$ is the most
sensitive parameter for the evolution of flux $S(\nu,R)$. To
illustrate this, we choose typical values of $p=2.2$, $\gamma_0=20$,
$R_0=4r_g$ and $v=0.004c$ and show the resulting model light curves
at 90 GHz while $B_0$ is 10, 20, 30, 40 and 50 Gauss in Figure 4.
Results from expanding plasmon model of van der Laan (1966) are
shown in dotted lines, and results from expanding plasmon model with
synchrotron cooling are shown in solid lines. The differences
between two results are significant above 30 Gauss, implying that
the synchrotron cooling of electrons should not be ignored for
strong magnetic field.

 The best-fit model for the light-curve of 2006 August 12 is
plotted as a solid line in Figure 5 left. Two blobs were required to
fit the data, which were assumed to appear at 7.0 and 10.3 UT. We
attribute the turnover in light curve to the birth of a new blob, so
the second blob was assumed to appear before the flux increases. The
corresponding initial blob radius $1.8^{+1.6}_{-0.6} r_g$ and
$2.6^{+0.4}_{-0.2} r_g$, expanding velocity $4^{+2}_{-1}\times
10^{-3} c$ and $2^{+2}_{-1}\times 10^{-3} c$, electron number
density of $2.56_{-2.52}\times 10^{7} {\rm cm}^{-3}$ and
$5.12^{+5.12}_{-2.56}\times 10^{7} {\rm cm}^{-3}$, magnetic field of
$19^{+30}_{-5}$ and $7^{+4}_{-2}$ Gauss were derived from the fit.
The uncertainty that was failed to be assessed was left blank, if
not such a sensitive parameter. The peak flux densities of two blobs
are estimated to be 0.26 and 0.59 Jy, respectively. The half-power
durations are 1.7 and 5.2 hr, respectively. Blob mass of $2.3\times
10^{20}$ g and $1.4\times 10^{21}$ g were estimated.

Figure 5 right shows the best-fit model for the light-curve of 2006
August 13. Similarly, three blobs appeared at 6.6, 10.0 and 13.0 UT
are required to fit the flare. Initial blob radius $4.2\pm0.2 r_g$,
$2.8^{+0.4}_{-0.6} r_g$ and $2.3^{+0.8}_{-0.4} r_g$, expanding
velocity $5^{+1}_{-2}\times 10^{-3} c$, $5^{+3}_{-1}\times 10^{-3}
c$ and $5\pm2\times 10^{-3} c$, electron number density of
$1.6^{+4.8}_{-1.2}\times 10^{6}{\rm cm}^{-3}$,
$2.56^{+7.68}_{-2.24}\times 10^{7} {\rm cm}^{-3}$ and
$5.12_{-4.48}\times 10^{7} {\rm cm}^{-3}$ and magnetic field of
$23^{+4}_{-8}$, $15^{+25}_{-8}$ and $13^{+8}_{-6}$ Gauss were
derived from the fit. The
 peak flux densities of three blobs are
0.67, 0.66 and 0.48 Jy, respectively. The half-power durations are
3.2, 2.8 and 2.4 hr, respectively. Blob mass of $1.8 \times 10^{20}$
g, $8.5\times 10^{20}$ g and $9.4\times 10^{20}$ g were estimated
based on derived parameters. The mass-loss rate contributed by blob
was then calculated to be $9.7\times10^{-10}M_{\odot}$ yr$^{-1}$.
This value is lower than the accretion rate range
$2\times10^{-7}M_{\odot}$ yr$^{-1}$ to $2\times10^{-9}M_{\odot}$
yr$^{-1}$ estimated by the rotation measure measurements (Marrone et
al. 2007). The derived parameters have been summarized in Table 2.

In principle, the expanding plasmon model can also be used to
interpret the 2000 March 7 NMA short millimeter flare reported by
Miyazaki et al. (2004). In their observation, the peak flux density
at the 140 GHz band is apparently larger than that at the 100 GHz
band. The spectral variation suggests that the energy injection to
photons occurred in the higher frequency regime first and the
emitting frequency was shifted to the millimeter-wavelength regime
with time, which is well consistent with the scenario predicted by
expanding plasmon model. A time delay of 1.5 hr was observed for NIR
and sub-mm flare on 2008 June 3 (Eckart et al. 2008b), which has
been explained with a similar model with adiabatically expanding
source components. There, the spectral index (0.9 to 1.8), expansion
velocity (0.005c) and source size ($\sim$2 $r_g$) are fairly
consistent with the parameters derived here. In order to compare
with their modeling results, we calculate the optical depth at
sub-mm based on parameters derived here. Take the first blob of 2006
Aug 13 for example, according to Pacholczyk (1970), the optical
depth at 90 GHz is calculated to be $8.23$ at $t_0$. The critical
optical depth at which the flux density for any particular frequency
peaks at $t_0$, is calculated to be $1.82$ based on Eq.(12).
According to Eq.(7), the optical depth at 345 GHz is $0.14$, which
is smaller than the critical value. Therefore, at both NIR and
sub-mm wavelengths, the emission is optically thin at the beginning.
Since we attribute the turnover in light curve to the emergence of a
new blob, so the new blob is assumed to appear right before the
observed flux density increases. Extending the model in time might
help give time delay between NIR and sub-mm, however, the birth time
of blob is difficult to determine. Future simultaneous
multi-wavelength observation, especially the correlation between
optically thin (such as NIR/X-ray) and 3-mm flaring emission is
expected to help improve the model fitting.

\subsection{The Hot Spot Model}

An orbiting hot spot model has been frequently used to mainly
explain the observations of short-term NIR and X-ray variability
(Broderick \& Loeb 2005, 2006, Meyer et al. 2006a, 2006b, Trippe et
al. 2007, Eckart et al. 2006b, 2008a). The hot spot is modeled by an
overdensity of non-thermal electrons centred at a certain point of
its Keplerian orbit. This situation may arise in the case of
magnetic reconnection event similar to the solar flare. Due to the
Doppler shift and relativistic beaming the approaching portion of
the hot spot orbit appears considerably brighter than the receding
portion. This model is successful in explaining the NIR 17 minutes
quasi-periodic oscillation (Genzel et al. 2003). The hot spot model
is applied to radio band by including the effects of disk opacity
for a typical RIAF model (Broderick \& Loeb 2006). In these studies,
the hot spot is always close to the innermost stable circular orbit
(ISCO), thus the NIR 17 minutes quasi-periodic oscillation can be
produced. Since the creation of such a hot spot is still under
discussion, it is also possible that such kind of spot may appear
somewhere away from the ISCO and thus produce quasi-periodic
oscillation with a longer timescale.

In the accretion disk, neighboring annuli of differentially rotating
matter experience a viscous shear that transports angular momentum
outwards and allows matter to slowly spiral in towards the center of
the potential (Merloni 2002). As a result, the gas rotates with a
sub-Keplerian angular velocity (Narayan et al. 1997). In the
following we assume that the rotation of hot spot is also
sub-Keplerian and fit our detected IDV events using a sub-Keplerian
rotating hot spot model. To simplify the calculation, the angular
velocity is assumed to be 0.4 times of Keplerian angular velocity of
a Schwarzschild black hole.

We assumed the values of most physical parameters of the hot spot
model the same as those in the expanding plasmon model when starting
fitting the hot spot model to light-curves. These include: the
energy range of relativistic particles, the particle spectral index
and the quiescent flux density. Then we estimated other parameters
by means of weighted least square method. The magnetic field was
assumed to range from 1 to 100 Gauss. In RIAF model, the electron
number density of accretion disk is about $2\times 10^{6} ~{\rm
cm}^{-3}$ at a distance $20 r_g$ from the central black hole (Yuan
et al. 2003). Since the hot spot is modeled by an overdensity of
non-thermal electrons, it is safe to assume the electron number
density ranges from $4\times 10^{6} $ to $1\times 10^{8}~{\rm
cm}^{-3}$. In addition, the accretion disk is assumed to be edge-on
to maximize the boosting effect (Huang et al. 2007; 2008). The final
result is the combination of the quiescent flux density and flux
density of hot spot. The derived parameters are summarized in Table
3.

The hot spot model for the 2006 August 12 flaring is plotted as a
solid line in Figure 6 left. The quiescent flux density was assumed
to be 1.4 Jy. Two hot spots are needed to fit the data. Radius of
$6.5\pm0.5 r_g$ and $8.0\pm0.5 r_g$, magnetic intensity of $3^{+2}$
and $1^{+2}$ Gauss and electron number density of $4^{+26}\times
10^{6}~{\rm cm}^{-3}$ and $4^{+26}\times 10^{6}~{\rm cm}^{-3}$ were
derived from the weighted least square fitting. The separation to
central black hole is $10\pm1 r_g$ and $12^{+2}_{-1} r_g$.

The hot spot model for the light-curve of 2006 August 13 is plotted
as a solid line in Figure 6 right. The quiescent flux density was
1.4 Jy too. One hot spot is required to fit the data. Radius of
$4.1\pm0.3 r_g$, magnetic intensity of $6^{+2}_{-5}$ Gauss and
electron number density of $6^{+44}\times 10^{6}~{\rm cm}^{-3}$ were
derived. The hot spot is at $11.4^{+0.4}_{-0.2}r_g$ from the central
black hole.

The electron cooling timescales due to synchrotron losses are
calculated to be greater than 2 days at 90 GHz, much longer than the
observed variation timescale, thus the synchrotron energy loss can
be ignored in the fitting. Since the synchrotron cooling time is
long, the life time of hot spot should mainly depend on the
dynamical timescale. Reid et al. (2008) analyzed the limits on the
position wander of Sgr A*, ruling out the possibility of hot spots
with orbital radius above $15 r_g$ that contribute more than 30\% of
the total 7-mm flux. All the orbital radius listed in Table 3 are
smaller than $15 r_g$. Hence, the presented hotspot model is not in
contradiction with their result.

The discussion above shows that both the expanding plasmon model and
the orbiting hot spot model can be used to interpret the detected
two IDV events. Given that the former model predicts a time delay in
flare emission, while the latter does not, the time delay between
different frequencies in flare emission is believed to be critical
to distinguish between them (Yusef-zadeh et al. 2006b). Recently,
Yusef-zadeh et al. (2008) detected time lags of 20.4$\pm$6.8,
30$\pm$12 and 20$\pm$6 minutes between the flare peaks observed at
13-mm and 7-mm. At shorter wavelength, a possible time delay of
110$\pm$17 minutes between X-rays and 850-$\mu$m was observed
(Marrone et al. 2008). Though these observations seem to support the
expanding plasmon model, the hot spot model is still a possible
explanation, especially for the observed nearly symmetrical
light-curves.

\section{SUMMARY}
\label{summary}


 We presented the results of the ATCA flux density monitoring of Sgr
A* at 3-mm, with emphasis on the detected two IDV events. Comparison
of flux densities in two observing sessions in 2006 indicates that
Sgr A* appeared to undergo a high state in June session, and a low
state in August session. On 2006 August 12, Sgr A* exhibits a 33\%
fractional variation in about 2.5 hr. Two peaks with a separation of
4 hr are seen on 2006 August 13 flare which exhibits a maximum
variation of 21\% within 2 hr.

The short timescales inspire us to consider mechanisms other than
synchrotron cooling that may be responsible for the variation. Both
the expanding plasmon model and the sub-Keplerian rotating hot spot
model were discussed and applied to interpret the observed light
curves. Because of a relatively large derived magnetic intensity
(and thus a short synchrotron cooling timescale), we incorporated
the synchrotron cooling into the original adiabatically expanding
plasmon model to model the observed IDV data. The radius of blob was
estimated to range from 1 to $5 r_{g}$, the expanding velocity range
from $0.001c$ to $0.007c$, the electron number density larger than
$1\times 10^{6}~{\rm cm}^{-3}$ and the magnetic field range from $7$
to $30$ Gauss. A minimum mass-loss rate of $ 9.7 \times
10^{-10}M_{\odot}$ yr$^{-1}$ was deduced based on these derived
parameters. We assume that the rotation of hot spot is sub-Keplerian
while applying the hotspot model. It seems that both models can
reasonably fit the detected IDV events. Future simultaneous
multi-wavelength monitoring is expected to discriminate them and
tell us where such kind of IDV events come from.

\acknowledgments
The Australia Telescope Compact Array is part of the Australia
Telescope which is founded by the Commonwealth of Australia for
operation as a National Facility managed by the CSIRO.

This work was supported in part by the National Natural Science
Foundation of China (grants 10573029, 10625314, 10633010 and
10821302) and the Knowledge Innovation Program of the Chinese
Academy of Sciences (Grant No. KJCX2-YW-T03), and sponsored by the
Program of Shanghai Subject Chief Scientist (06XD14024) and the
National Key Basic Research Development Program of China (No.
2007CB815405).


\begin{figure}[]
\begin{center}
\includegraphics[width=3.6in]{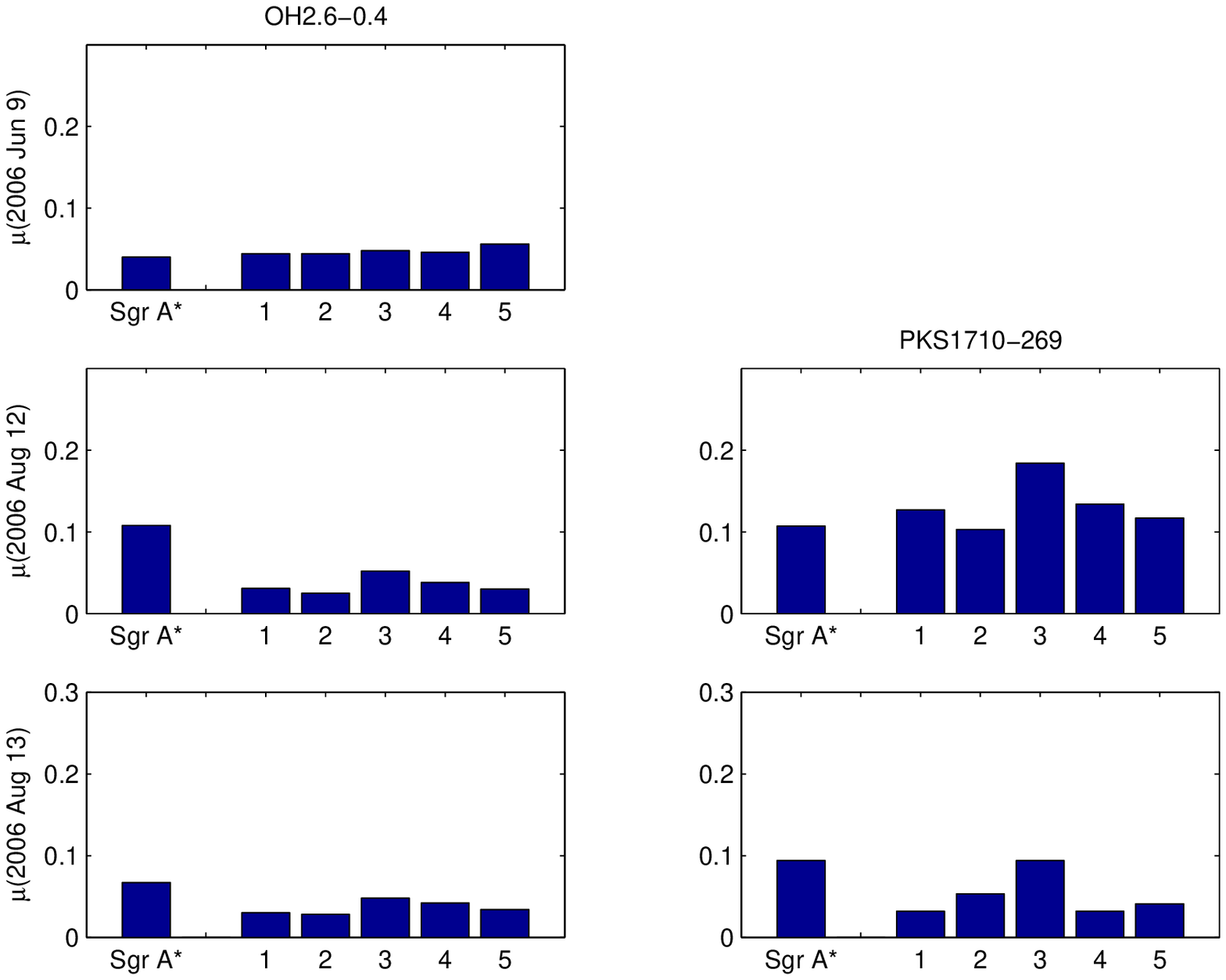}
\vspace*{6 mm} \caption{ Modulation index of flux density of Sgr A*,
and gain corrections of five antennas (labeled 1, 2, 3, 4 and 5).
They are derived from calibrators OH2.6-0.4 and PKS 1710-269 (from
left to right) for three observations on 2006 June 9, August 12 and
13 (from top to bottom). Many data obtained from antenna 3 in August
session were shadowed and have been flagged, so the uncertainty of
this antenna is big compared with other antennas. During observation
in 2006 August, the 3-mm flux density of PKS 1710-269 was around
0.5~Jy, only one fiftieth of that of OH2.6-0.4, therefore the
modulation index of gain corrections derived from this source are
particularly high.}
   \label{fig1}
\end{center}
\end{figure}

\clearpage

\begin{figure}[]
\begin{center}
\includegraphics[width=6.6in]{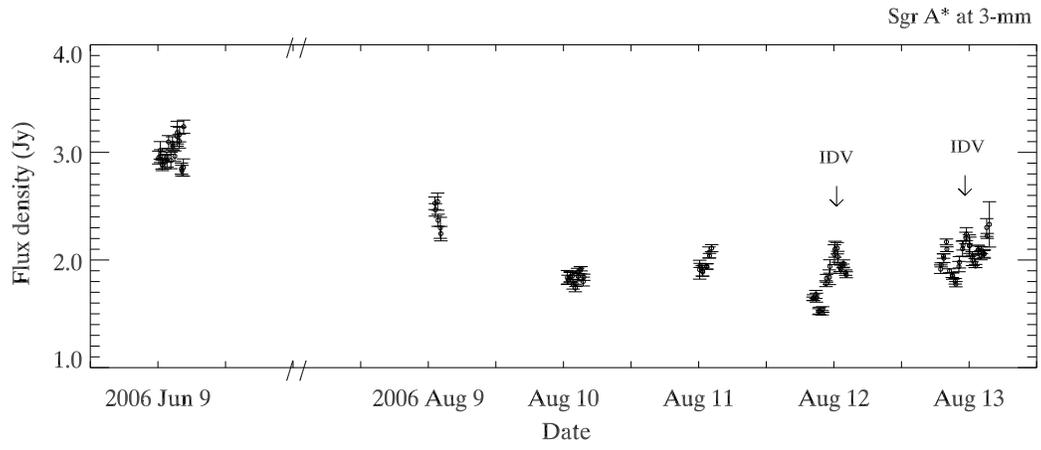}
\vspace*{6 mm} \caption{ ATCA 3-mm light-curves of Sgr A* in 2006.
Two detected IDV events are indicated (arrows).}
   \label{fig1}
\end{center}
\end{figure}

\clearpage

\begin{figure}[]
\begin{center}
\includegraphics[width=3.2in]{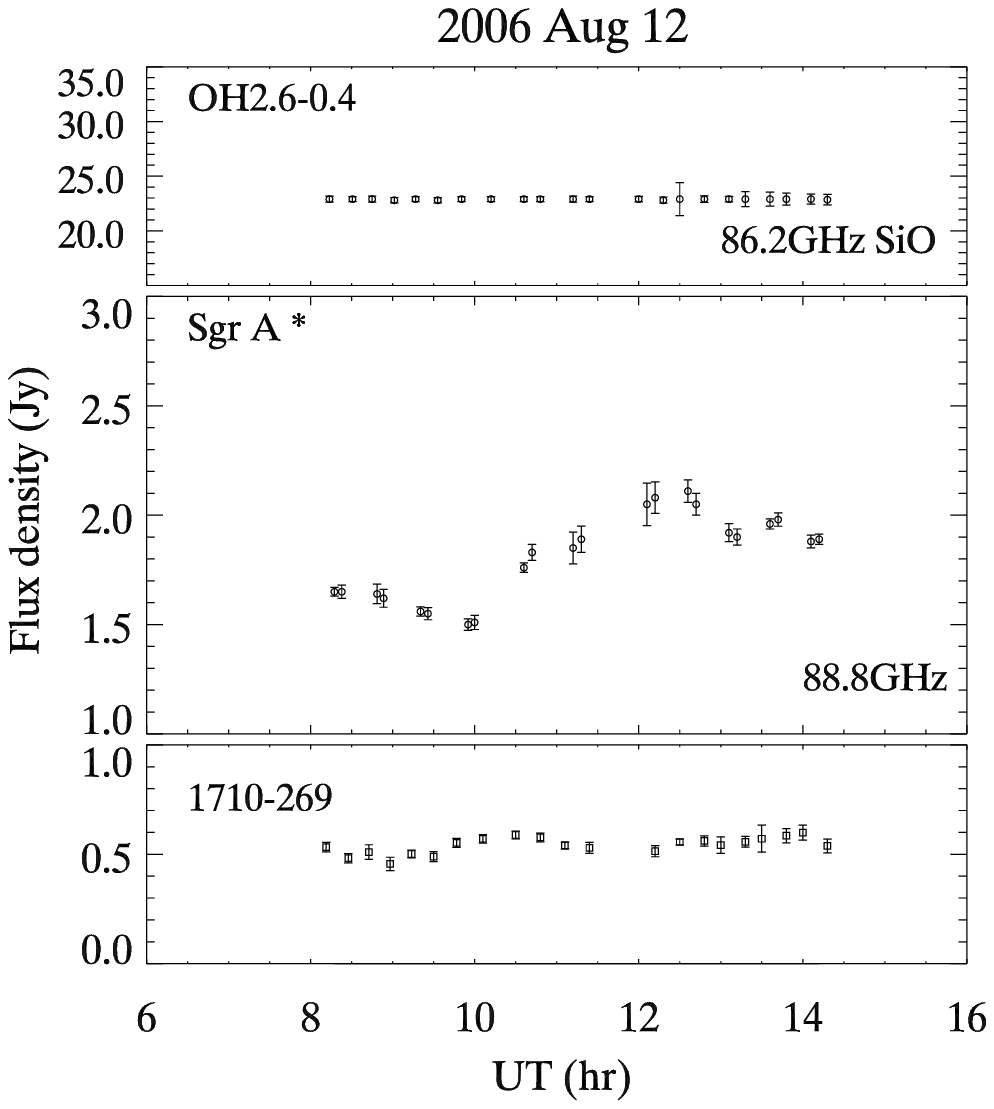}
\includegraphics[width=3.2in]{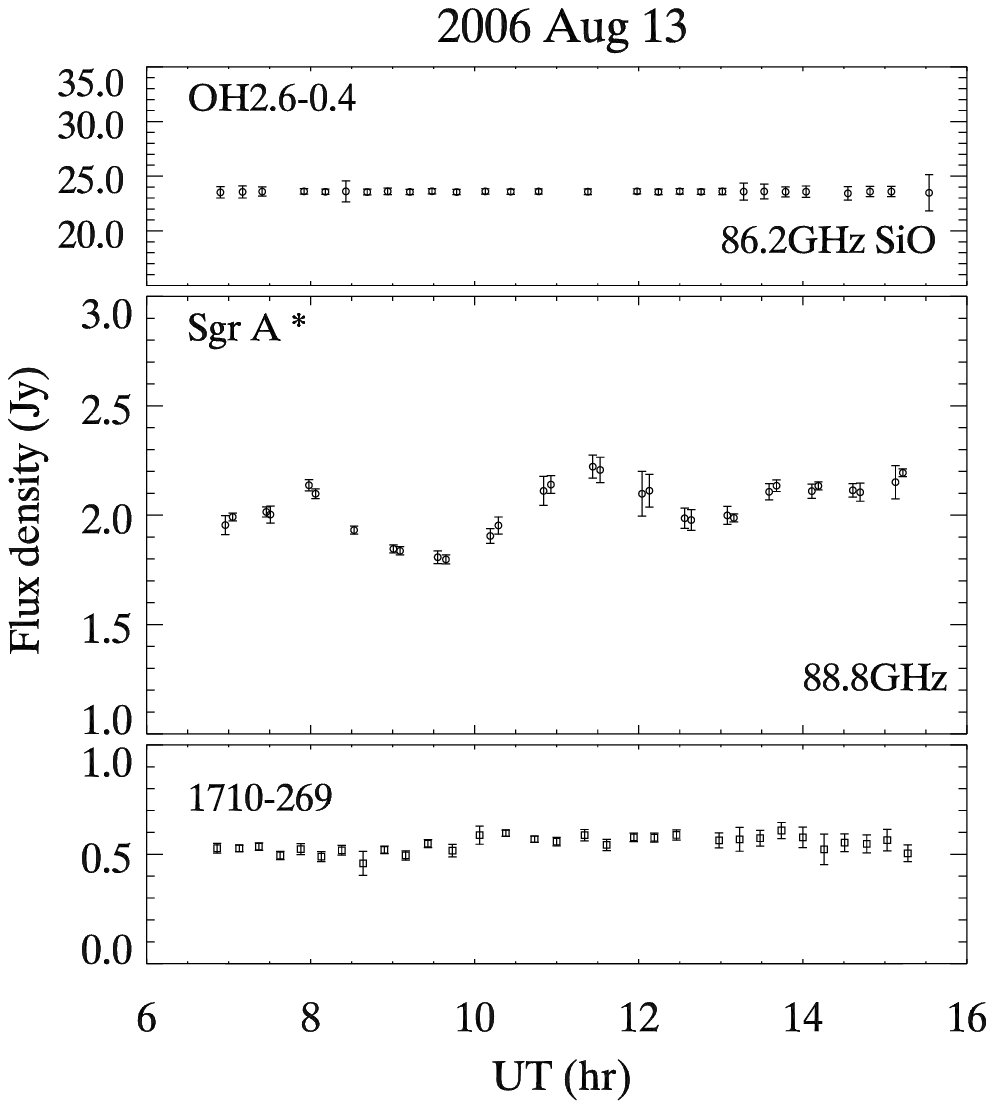}
\vspace*{-0.2 cm} \caption{ATCA 3-mm light-curves on 2006 August 12
(Left) and 13 (Right) of Sgr A* (middle panel), secondary
calibrators OH 2.6-0.4 (top panel) and PKS 1710-269 (bottom panel).}
   \label{fig2}
\end{center}
\end{figure}

\clearpage

\begin{figure}[]
\begin{center}
\includegraphics[width=4.6in]{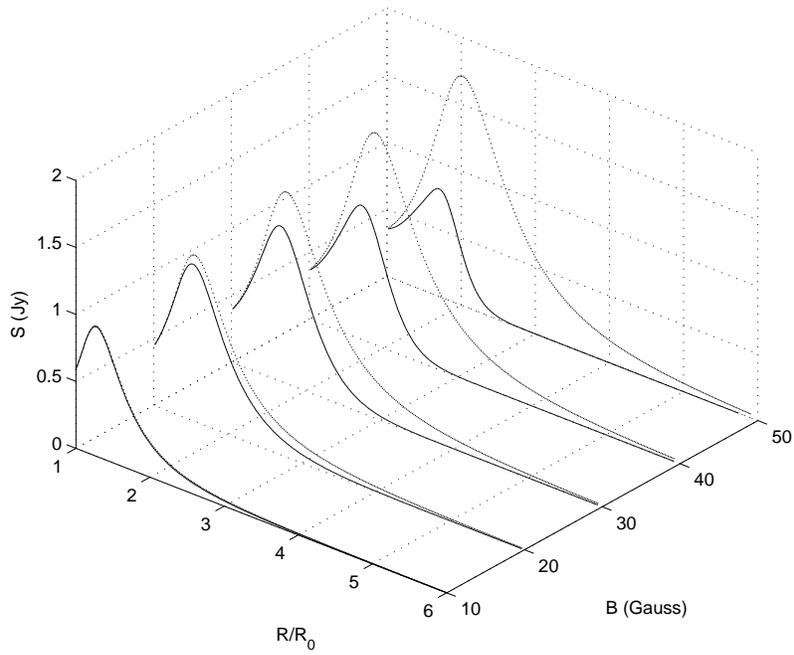}
\vspace*{6 mm} \caption{Two kind of theoretical model light curves
as a function of expanding blob radius at 90 GHz with different
$B_0$. Results from expanding plasmon model of van der Laan (1966)
are shown in dotted lines, and results from expanding plasmon model
with synchrotron cooling are shown in solid lines. }
   \label{fig1}
\end{center}
\end{figure}

\clearpage

\begin{figure}[]
\begin{center}
\includegraphics[width=3.2in]{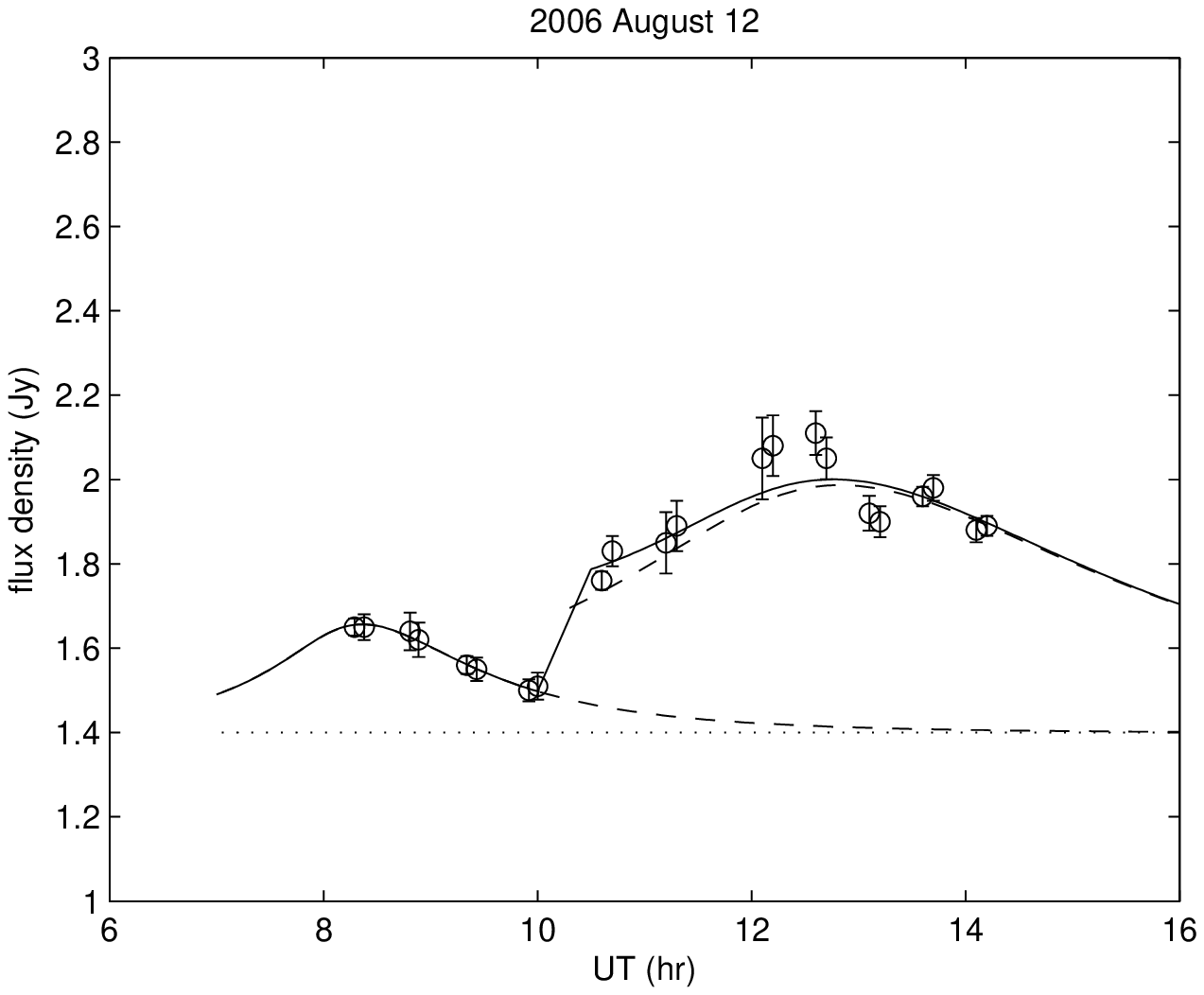}
\includegraphics[width=3.2in]{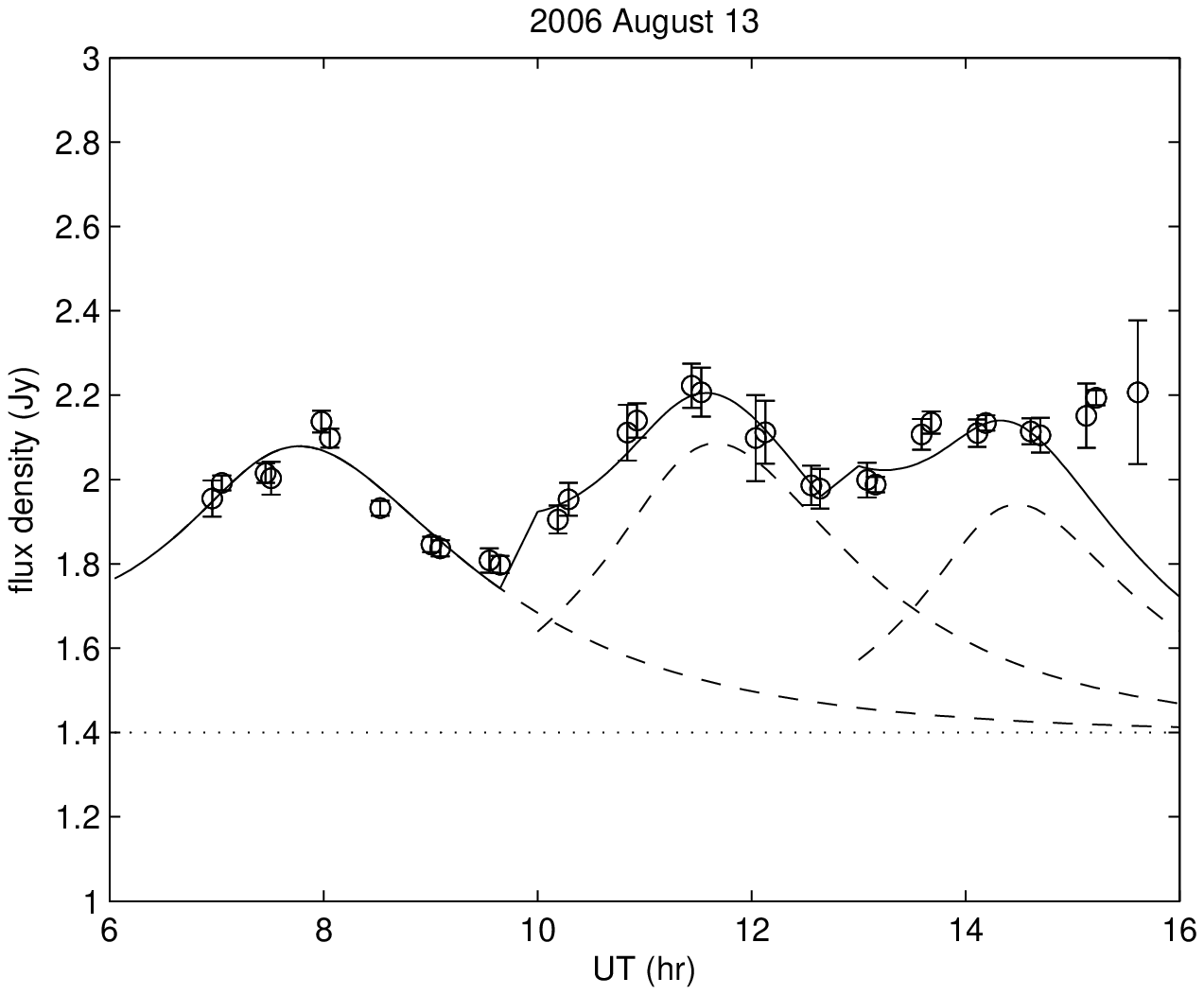}
\vspace*{-0.2 cm} \caption{The solid line represents the expanding
plasmon model fitting to the observed 3-mm light-curves on 2006
August 12 (left) and 13 (right) with synchrotron radiation cooling
taken into account. An assumed quiescent flux density of 1.4 Jy is
indicated by the straight dotted line. The blobs used to fit the
data are indicated by dashed curves.}
   \label{fig3}
\end{center}
\end{figure}

\clearpage

\begin{figure}[]
\begin{center}
\includegraphics[width=3.2in]{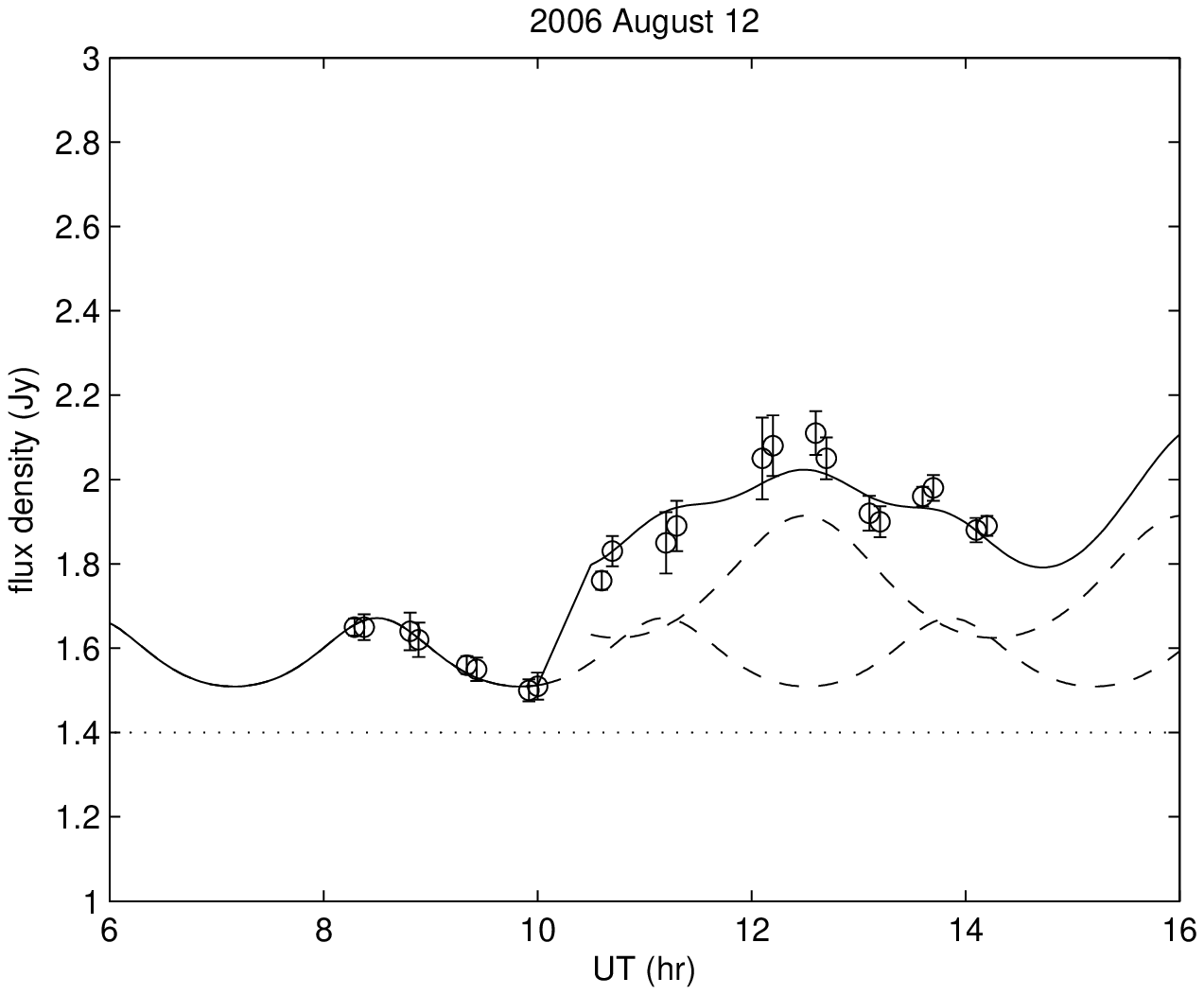}
\includegraphics[width=3.2in]{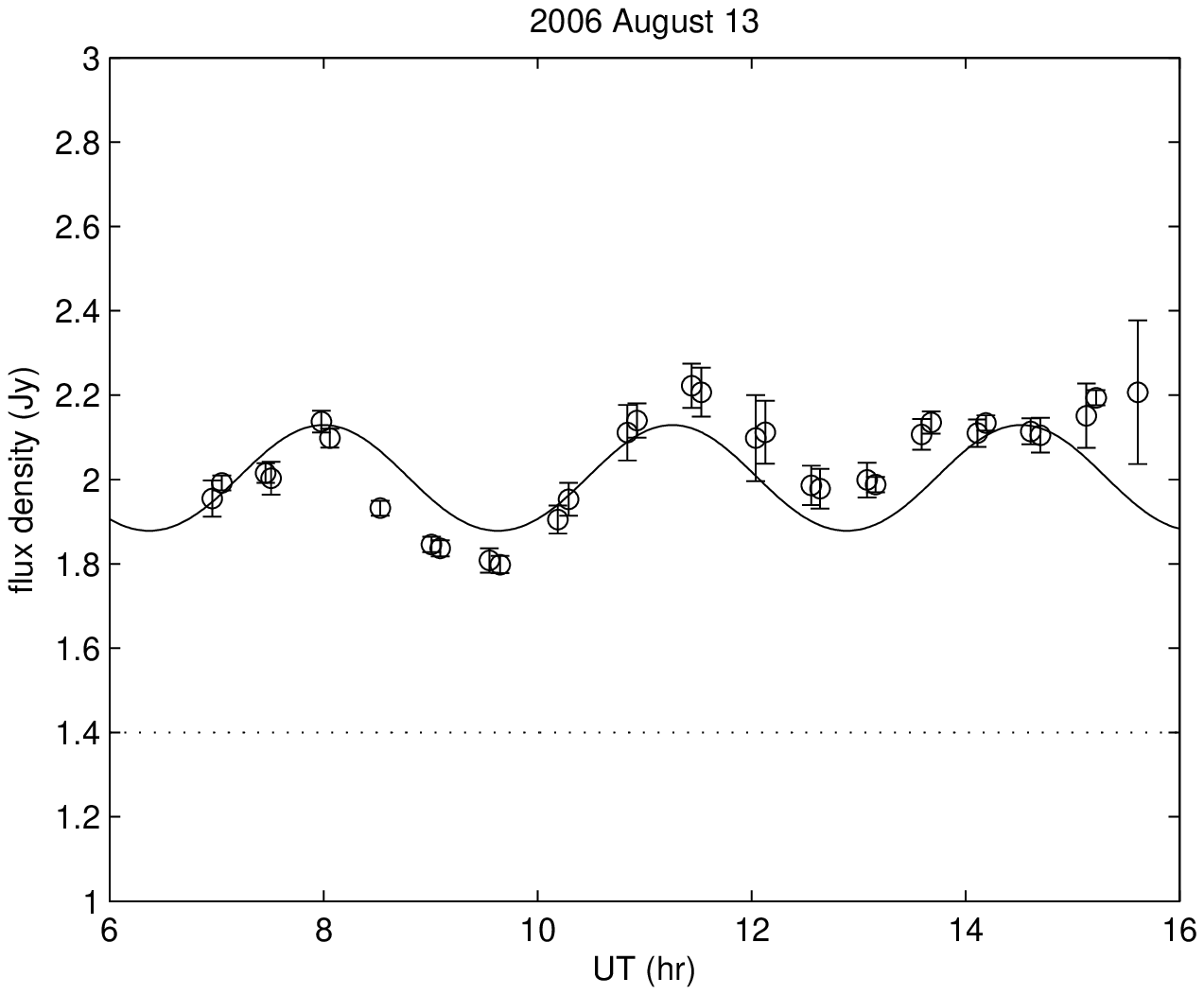}
\vspace*{-0.2 cm} \caption{The light-curves produced by the
sub-Keplerian orbiting hot spot model on 2006 August 12 (left) and
13 (right). An assumed quiescent flux density of 1.4 Jy is indicated
by the straight dotted line. Two hot spots used to fit the data on
August 12 (left) are indicated by dashed curves.}
   \label{fig4}
\end{center}
\end{figure}

\clearpage

\begin{table}
    \begin{center}
      \caption{ ATCA Observations of Sgr A* in 2006. Length is duration of the observation.
IF1\&IF2 are intermediate frequencies for the lower and upper
sidebands, respectively, with the corresponding bandwidth of
BW1\&BW2. Range of baselines are indicated by uv range. Beam is the
ATCA synthesized beam. }\label{tab:fractionalvariation}
      \begin{tabular}{lccclccccccc}
    \hline
 Date & Length  & IF1\&IF2 & BW1\&BW2 &  uv range & Beam   \\
      &  (hr)   &    (GHz) & (MHz) &  (k$\lambda$)   &   &  \\
            \hline
2005 Oct 18 & 10  & 93.504\&95.552 & 128\&128  & 13-61    &  $2.^{\prime\prime}9 \times 1.^{\prime\prime}7$  \\
            \hline
2006 June 9 & 6  & 86.243\&88.896 & 16\&128  &  20-430  & $2.^{\prime\prime}1 \times 0.^{\prime\prime}3$  \\
\hline
2006 Aug 9  &  1 &                    &             &     &                          \\
2006 Aug 10 & 4  &                     &            &       &                                             \\
2006 Aug 11 &  2 &   86.243\&88.896    & 16\&128    &  3-570  & $1.^{\prime\prime}3 \times  0.^{\prime\prime}2$ \\
2006 Aug 12 &  7  &                     &        &        &                                         \\
2006 Aug 13  & 9   &                    &        &         &                                       \\
            \hline
      \end{tabular}

  \end{center}
\end{table}

\begin{table}
    \begin{center}
      \caption{Parameters of the expanding plasmon model. We took into account the synchrotron cooling of
electrons while applying the adiabatically expanding plasmon model
to the light-curves. Here, $t_{0}$ is the time at which the blob was
assumed to be generated, $R_{0}$ is the initial radius, $v$ is the
expanding velocity, $N_{0}$ is the initial electron number density,
$B_{0}$ is the initial magnetic field strength, $S_{\rm p}$ is the
peak flux density of the blob, and $\chi^{2}_{\nu}$ is the reduced
chi squares. }\label{tab:parameters}
      \begin{tabular}{lcccccccccc}
    \hline
  Date & $t_{0}$(hr)& $R_{0}$($r_{g}$)& v ($10^{-3}$c)& $N_{0}$ (${\rm cm}^{-3}$) & $B_{0}$(Gauss) & $S_{\rm p}$(mJy) & $\chi^{2}_{\nu}$ \\
\hline
 2006 Aug 12 & 7.0 &  $1.8^{+1.6}_{-0.6}$ & $4^{+2}_{-1}$  & $2.56_{-2.52}\times 10^{7}$  & $19^{+30}_{-5}$  &  0.26 & 1.64 \\
             & 10.5 &  $2.6^{+0.4}_{-0.2}$  & $2^{+2}_{-1}$  & $5.12^{+5.12}_{-2.56}\times 10^{7}$  & $7^{+4}_{-2}$  &  0.59 &     \\
              \hline
 2006 Aug 13 & 6.6 & $4.2\pm0.2$ & 5$^{+1}_{-2}$  &  $1.6^{+4.8}_{-1.2}\times 10^{6}$  & $23^{+4}_{-8}$  &  0.67  & 3.60 \\
             & 10.0 & $2.8^{+0.4}_{-0.6}$ & $5^{+3}_{-1}$ & $2.56^{+7.68}_{-2.24}\times 10^{7}$  & $15^{+25}_{-8}$  &  0.66 &     \\
            & 13.0 &  $2.3^{+0.8}_{-0.4}$  & $5\pm2$  & $5.12_{-4.48}\times 10^{7}$  & $13^{+8}_{-6}$ & 0.48 &   \\
            \hline
      \end{tabular}

  \end{center}
\end{table}

\begin{table}
    \begin{center}
      \caption{Parameters of the sub-Keplerian orbiting hot spot model. $R$ is the radius
      of hot spot, $N_{e}$ is the electron number density, B is
the magnetic field strength, D is the distance between hot spot and
the central black hole, and $\chi^{2}_{\nu}$ is the reduced chi
squares.}\label{tab:parameters2}
      \begin{tabular}{lccccccccc}
    \hline

 Date & $R$ ($r_{g}$)  & $N_{e}$ (${\rm cm}^{-3}$) & $B$ (Gauss) & $D$ ($r_{g}$) & $\chi^{2}_{\nu}$ \\
\hline
     2006 Aug 12    & $6.5\pm0.5$    &  $4^{+26}\times 10^{6}$ & $3^{+2}$  & $10\pm1$  & 1.04  \\
   & $8.0\pm0.5$    &  $4^{+26}\times 10^{6}$ & $1^{+2}$
   &  $12^{+2}_{-1}$ &  \\
\hline
     2006 Aug 13    &  $4.1\pm0.3$  &  $6^{+44}\times 10^{6}$ & $6^{+2}_{-5}$  & $11.4^{+0.4}_{-0.2}$  &  12.32  \\
            \hline
      \end{tabular}

  \end{center}
\end{table}

\end{document}